\documentclass[useAMS,usenatbib]{mn2e}
\usepackage{txfonts}
\usepackage{natbib}
\usepackage[all]{xy}

\usepackage[dvips]{graphicx}

\def\del#1{{}}

\newcommand{\ltsima}{$\; \buildrel < \over \sim \;$}
\newcommand{\lsim}{\lower.5ex\hbox{\ltsima}}
\newcommand{\gtsima}{$\; \buildrel > \over \sim \;$}
\newcommand{\gsim}{\lower.5ex\hbox{\gtsima}}

\newcommand{\dd}{\mathrm{d}}

\newcommand{\ttt}{\Theta\Theta}
\newcommand{\cgl}{C_{\mathrm{gl}}}
\newcommand{\cgltwo}{C_{\mathrm{gl,2}}}
\newcommand{\eqref}[1]{(\ref{#1})}
\newcommand{\dax}[1]{\left\langle \delta\alpha_x^{#1}\right\rangle}
\newcommand{\dayy}[1]{\left\langle \delta\alpha_y^{#1}\right\rangle}

\title[CMB lensing by nonlinear structures]
{Gravitational lensing of the cosmic microwave background by nonlinear structures}
\author[Philipp M. Merkel and Bj{\"o}rn Malte Sch\"afer]
{Philipp M. Merkel\thanks{e-mail: Philipp.Merkel@ita.uni-heidelberg.de} and Bj{\"o}rn Malte Sch\"afer\\
Institut f{\"u}r Theoretische Astrophysik, Zentrum f{\"u}r Astronomie, Universit{\"a}t Heidelberg, Albert-Ueberle-Stra{\ss}e 2, 69120 Heidelberg, Germany\\
Astronomisches Recheninstitut, Zentrum f{\"u}r Astronomie, Universit{\"a}t Heidelberg, M{\"o}nchhofstra{\ss}e 12, 69120 Heidelberg, Germany}

\begin{document}
\onecolumn
\pagerange{\pageref{firstpage}--\pageref{lastpage}}
\pubyear{2010}
\maketitle
\label{firstpage}

\begin{abstract}
Weak gravitational lensing changes the angular power spectra of the cosmic microwave background (CMB) temperature and polarization in a characteristic way containing valuable information for cosmological parameter estimation and weak lensing reconstructions. So far, analytical expressions for the lensed CMB power spectra assume the probability density function (PDF) of the lensing excursion angle to be Gaussian. However, coherent light deflection by nonlinear structures at low redshifts causes deviations from a pure Gaussian PDF. Working in the flat-sky limit we develop a method for computing the lensed CMB power spectra which takes these non-Gaussian features into account.
Our method does not assume any specific PDF but uses instead an expansion of the characteristic function of the lensing excursion angle into its moments.
Measuring these in the CMB lensing deflection field obtained from the Millennium Simulation we show that the change in the lensed power spectra is only at the 0.1\%--0.4\% level on very small scales \((\Delta \theta \lesssim 4', \, l\gtrsim 2500)\) and demonstrate that the assumption of a Gaussian lensing excursion angle PDF is well applicable.
\end{abstract}


\section{Introduction}
Cosmological parameter analysis as well as investigation of inflationary models require a precise knowledge of the fluctuations of the cosmic microwave background. Therefore supreme technical efforts are made to equip current and future CMB experiments with instruments with ever greater resolution including sensitivity on CMB polarization.
In order to interpret these low-noise data correctly a profound understanding of the physics of the primary CMB is not longer sufficient but requires an equivally good understanding of CMB foregrounds.

One of the most important of these is weak gravitational lensing: on their way from the last-scattering-surface to today's observer CMB photons are deflected by the intervening large-scale structure. Although each single deflection is small, their cumulative effect changes the statistics of the CMB fluctuations and accordingly their power spectra observed on today's sky \citep[see][]{2001PhR...340..291B, 2006PhR...429....1L}. At present, the lensing effect has been detected in the temperature data of the WMAP satellite at moderate significance \citep{2007PhRvD..76d3510S, 2008PhRvD..78d3520H}. Apart from being a CMB foreground, the lensing pattern itself is a valuable source of cosmological information \citep{2000PhRvD..62d3517G, 2002ApJ...574..566H, 2003PhRvD..67d3001H, 2003PhRvD..67h3002O, 2004NewA....9..687A, 2006PhRvD..73d5021L, 2006JCAP...10..013P, 2009PhRvD..79f5033D}.

\citet{1996ApJ...463....1S,1998PhRvD..58b3003Z,2005PhRvD..71j3010C} developed efficient and accurate methods for computing the lensed CMB power spectra starting from the corresponding correlation functions. These methods all work under the assumption of a Gaussian distributed lensing excursion angle, which measures the difference between the deflection of two nearby light rays. Since the deflection angle is given by the gradient of the lensing potential one commonly accounts for the impact of nonlinear structures on the lensed CMB spectra semi-analytically by applying nonlinear but Gaussian corrections from the HALOFIT model of \citet{2003MNRAS.341.1311S} to the lensing potential power spectrum. This approach takes the additional power on small scales due to nonlinear evolution into account but ignores that coherent deflection by nonlinear structures at low redshifts alters the Gaussian character of the lensing excursion angle distribution. In this work we investigate how exactly this non-Gaussianity influences the lensed CMB power spectra.

This paper is structured as follows. In Section~\ref{sec:formalism} we develop the formalism for computing the lensed CMB temperature power spectrum in case of a non-Gaussian lensing excursion angle PDF. We relegate the discussion of the power spectra involving CMB polarization to Appendix~\ref{sec:polarization}. Section~\ref{sec:numerics} is devoted to our numerical results. Here we quantify the non-Gaussianity present in the lensing excursion angle PDF which we compute from the CMB deflection field of the Millennium Simulation (MS). With this deflection field we then derive the lensed CMB power spectra and compare our results to power spectra resulting from using a pure Gaussian lensing excursion angle PDF. In Section~\ref{sect_summary} we summarize our results and give an outlook on future investigations. Finally, in Appendix~\ref{sec:explicit_expressions} we give explicit expressions for the lensed CMB correlation functions used in our numerical computations.

Throughout this work we assume a spatially flat \(\Lambda\textnormal{CDM}\) cosmology with adiabatic Gaussian initial perturbations. The relevant parameter values are: \(\Omega_m=0.25\), \(\Omega_\Lambda=0.75\), \(H_0 = 73 \, \textnormal{km} \, \textnormal{s}^{-1} \, \textnormal{Mpc}^{-1}\), \(\Omega_b=0.045\), \(n_s=1.0\), \(\sigma_8 = 0.9\) and \(r = 0.0\) (no primordial gravitational waves present). These parameters are equal to those of the Millennium Simulation \citep{2006Natur.440.1137S}.

\section{Formalism}\label{sec:formalism}

\subsection{Lensed CMB temperature power spectrum}\label{subsec:lensed_CMB_temp_power_spectrum}

Working in the flat-sky limit the 2D lensed temperature field is given by the remapping 
\begin{equation}
 \tilde{\Theta} ( \mathbf x ) = \Theta ( \mathbf x + \balpha (\mathbf x ) )
\end{equation}
mediated by the lensing deflection angle \(\balpha\), i.e. the gradient of the lensing potential \(\psi\): \(\balpha = \nabla \psi\).
In the approximation of instantaneous recombination the CMB can be described by a single source plane at conformal distance \(\chi = \chi^*\). In the absence of anisotropic stress the lensing potential is then given by the line of sight projection of the physical peculiar gravitational potential \(\phi\):
\begin{equation}
	\psi \left(\mathbf{\hat{n}}\right) = \frac{2}{c^2} \int\limits_{0}^{\chi^*} \dd \chi \;
	\frac{\chi^* - \chi}{\chi^*\chi} \phi \left( \chi \mathbf{\hat{n}} ; \chi \right) ,
\end{equation}
where the underlying geometry is flat \citep{2001PhR...340..291B, 2006PhR...429....1L}.
Introducing the Fourier transform of the temperature field via
\begin{equation}
 \Theta(\mathbf x ) = \int \frac{\dd^2 l}{2\pi} \Theta (\mathbf l ) e^{\mathrm i \mathbf l \cdot \mathbf x},
\end{equation}
the spectrum for a statistically homogeneous and isotropic field reads
\begin{equation}
 \left\langle \Theta (\mathbf l) \Theta^* ( \mathbf l' ) \right\rangle = C_l^{\ttt} \delta ( \mathbf l - \mathbf l' ).
\end{equation}
Then, ignoring the weak large-scale correlation between CMB and lensing potential due to the integrated Sachs-Wolfe effect, the lensed correlation function of the CMB temperature fluctuations is given by
\begin{equation}
 \tilde{\xi} ( r ) 	= \left\langle \tilde{\Theta} ( \mathbf x ) \tilde{\Theta} ( \mathbf x' ) \right\rangle
			= \int \frac{\dd^2 l}{(2\pi)^2} C_l^{\ttt} 
				e^{\mathrm i \mathbf l \cdot \mathbf x} 
				\left\langle e^{\mathrm i \mathbf l \cdot \left[ \balpha (\mathbf x) - \balpha (\mathbf x')\right]} \right\rangle
 \label{eq:lensed_correlation_function}
\end{equation}
where \(r = \left|\mathbf x - \mathbf x'\right|\).
It is worth noting that the lensed correlation function only depends on the \emph{relative} displacement, the so-called lensing excursion angle, \(\bdelta\balpha(\mathbf r) \equiv \balpha (\mathbf x) - \balpha (\mathbf x')\) and that this dependence is given by the \emph{characteristic function} of the lensing excursion angle:
\begin{equation}
 	\varphi_{\bdelta\balpha} (\mathbf l ) \equiv \left\langle e^{\mathrm i \mathbf l \cdot \mathbf \bdelta \balpha} \right\rangle =
	\int \dd \left(\bdelta\balpha\right) \; p\left(\bdelta\balpha\right) e^{\mathrm i \mathbf l \cdot \mathbf \bdelta \balpha}.
\end{equation}
The last equality reveals that the characteristic function is the Fourier transform of the PDF \(p\left(\bdelta \balpha\right)\). Hence, it carries the same information as the PDF itself.
From equation~\eqref{eq:lensed_correlation_function} the lensed power spectrum is readily obtained by
\begin{equation}
 \tilde C_l^{\ttt} = 2\pi \int \dd r \; r \tilde \xi (r) J_0(lr)
\end{equation}
where \(J_n(z)\) denotes the \(n\)-th order Bessel function \citep{1965hmfw.book.....A}.

\subsection{Lensing excursion angle}\label{subsec:lensing_ex_angle}

In linear theory the lensing potential is a Gaussian field and so is its gradient, the lensing deflection angle. Accordingly, the lensing excursion angle is a Gaussian variate and therefore the characteristic function of the lensing excursion angle is just given in terms of the variance
\begin{equation}
   \left\langle e^{\mathrm i \mathbf l \cdot \left[ \balpha (\mathbf x) - \balpha (\mathbf x')\right]} \right\rangle 
	 = \exp \left( -\frac{1}{2}\left\langle \left[ \mathbf l \cdot \bdelta\balpha \right]^2\right\rangle  \right)
	 = \exp \left( -\frac{1}{2} l^2 \left[\sigma^2(r) + \cos2(\phi_l - \phi_r) \cgltwo(r) \right] \right)
 \label{eq:gaussian_char_fct}
\end{equation}
where \(\phi_{\mathbf l, \mathbf r}\) denotes the angle between \(\mathbf l\), \(\mathbf r\) and the \(x\)-axis. We have defined \(\sigma^2(r) = \frac{1}{2}\bigl\langle \bdelta\balpha^2 \bigr\rangle = \frac{1}{2}\left(\cgl(0) - \cgl(r)\right)\). \(\cgl\) and \(\cgltwo\) are given in terms of the power spectrum of the lensing potential \(C^{\psi\psi}_l\) \citep{2005PhRvD..71j3010C}:
\begin{equation}
 \cgl(r) = \frac{1}{2\pi} \int \dd l \; l^3 C_l^{\psi\psi} J_0(lr) \qquad \textnormal{and} \qquad \cgltwo(r) = \frac{1}{2\pi} \int \dd l \; l^3 C_l^{\psi\psi} J_2(lr).
 \label{eq:def_of_cgl_and_cgl2}
\end{equation}

Inserting equation~\eqref{eq:def_of_cgl_and_cgl2} into equation~\eqref{eq:lensed_correlation_function} and performing a perturbative expansion in \(\cgltwo\) up to second order one recovers the expressions derived by \citet{2005PhRvD..71j3010C}.
However, \citet{2001MNRAS.327..169H} and \citet{2005MNRAS.356..829H} showed in numerical weak lensing ray-tracing experiments that the lensing excursion angle is not Gaussian distributed. 
Its PDF has indeed a Gaussian core but also exponential wings. These wings are a consequence of coherent scattering by individual massive haloes with mass larger than \(10^{14}M_{\sun}/h\). Since coherent deflection demands a sufficiently small intrinsic separation of the light rays the exponential wings appear prominent in the excursion angle PDFs obtained from light rays with intrinsic separation of a few arcminutes and are negligible in those of separations larger than one degree \citep{2005MNRAS.356..829H}. The contributions from coherent scattering broaden the PDFs, i.e. larger excursion angles are more probable than in case of a pure Gaussian PDF. For example, for intrinsic separations smaller than two arcminutes an excursion angle of one arcminute is about ten times more probable than for a Gaussian PDF.

\subsection{Non-Gaussian probability density function}\label{subsec:non_gaussian_pdf}

In order to investigate how the non-Gaussian features of the lensing excursion angle PDF affects the lensed CMB temperature spectrum one has to use (in principle) all moments of the lensing excursion angle for computing the lensed correlation function (\ref{eq:lensed_correlation_function}), which is now denoted with a hat to distinguish it from the correlation function derived under the Gaussian assumption (denoted with a tilde)
\begin{equation}
 \hat\xi(r) 	= \left\langle \tilde \Theta (\mathbf x ) \tilde \Theta(\mathbf x') \right\rangle 
		= \int \frac{\dd^2l}{(2\pi)^2} C_l^{\ttt} e^{\mathrm i \mathbf l \cdot \mathbf r} 
			\left\langle \sum_{n=0}^{\infty} \frac{\left( \mathrm i \mathbf l \cdot \bdelta \balpha\right)^n}{n!}\right\rangle
		= \int \frac{\dd^2l}{(2\pi)^2} C_l^{\ttt} e^{\mathrm i \mathbf l \cdot \mathbf r}
			\sum_{n=0}^{\infty} \frac{\mathrm i^n}{n!} \sum_{k=0}^{n} {n \choose k} l_x^kl_y^{n-k} 
			\left\langle \delta\alpha_x^k \delta\alpha_y^{n-k}\right\rangle.
\end{equation}
For the last equality we used the binomial law
\begin{equation}
	\left( \mathbf l \cdot \bdelta\balpha \right)^n = \left( l_x\delta\alpha_x + l_y \delta\alpha_y \right)^n 
	= \sum_{k=0}^{n} {n \choose k} \bigl(l_x \delta\alpha_x\bigr)^k \bigl(l_y\delta\alpha_y\bigr)^{n-k}
	\quad \textnormal{with} \quad {n \choose k} = \frac{n!}{k!\, (n-k)!}.
\end{equation}
Following \citet{1997PhRvD..55.7368K} one should choose the local coordinate system, in which to define the correlation function, aligned with the great circle connecting the two points where the temperature fluctuations are measured. In the flat-sky limit this choice of coordinate system corresponds to evaluating the two-point correlator at the origin of the flat coordinate system and at a point on the \(x\)-axis at distance \(r\). One of the great advantages of this coordinate system is that here the correlation tensor of the lensing excursion angle is diagonal, i.e \(\left\langle \alpha_i (\mathbf x) \alpha_{j}(\mathbf x') \right\rangle \propto \delta_{ij}\). In case of a Gaussian distribution it then follows immediately by virtue of Wick's theorem that different moments of different components are statistically independent. In Section~\ref{subsec:corr_coeff} we will show that it is reasonable to assign this property also to the (non-Gaussian) PDF of the lensing excursion angle. Hence, introducing polar coordinates \(\mathbf l = (l\cos\phi, l\sin\phi)\), the lensed correlation function reads
\begin{equation}
 \hat\xi(r)	= \frac{1}{(2\pi)^2} \int \int \dd\phi \: l\dd l \; C^{\ttt}_l e^{\mathrm i l r \cos \phi}
			\sum_{n=0}^{\infty} \frac{\mathrm i^n}{n!} \sum_{k=0}^{n} {n \choose k} l^n\cos^k\phi \sin^{n-k}\phi
			\left\langle \delta\alpha_x^k \right\rangle \left\langle \delta\alpha_y^{n-k}\right\rangle.
\end{equation}
A further simplification can be obtained by demanding that components of the lensing excursion angle are distributed symmetrically about zero. This assumption is very natural since otherwise there would be a preferred direction along the coordinate axes. For a symmetric PDF all odd moments vanish, hence, 
\begin{eqnarray}
	\hat\xi(r)
	& = & 
	\frac{1}{(2\pi)^2} \int \int \dd\phi \: l\dd l \; C^{\ttt}_l e^{\mathrm i l r \cos \phi}
	\sum_{n=0}^\infty \sum_{k=0}^{n} \sum_{q=0}^{n-k} \frac{(-1)^{n+q}}{(2n)!}  {2n \choose 2k} {n-k \choose q}l^{2n} \cos^{2(k+q)} \phi
	\left\langle \delta\alpha_x^{2k} \right\rangle \left\langle \delta\alpha_y^{2(n-k)}\right\rangle
 	\nonumber\\
 	& = &
	\frac{1}{2\pi} \int l\dd l \; C^{\ttt}_l
	\sum_{n=0}^\infty \sum_{k=0}^{n} \sum_{q=0}^{n-k} \sum_{r=0}^{2(k+q)}
	\frac{(-1)^{n+k+r}}{(2n)!4^{k+q}}  {2n \choose 2k} {n-k \choose q} {2(k+q) \choose r}
	l^{2n} J_{2(k+q-r)}(lr)
	\left\langle \delta\alpha_x^{2k} \right\rangle \left\langle \delta\alpha_y^{2(n-k)}\right\rangle.
 \label{eq:lensed_corr_fct_non_linear}
\end{eqnarray}
To get the second line we used the fact that
\begin{equation}
 \int \dd \phi \; e^{\mathrm i z \cos \phi} \cos^n \phi = 2\pi (-\mathrm i )^n \frac{\dd^n}{\dd z^n} \: J_0(z)
\end{equation}
and that
\begin{equation}
 \frac{\dd^n}{\dd z^n} \: J_0 (z) = \frac{1}{2^n} \sum_{k=0}^n(-1)^k { n \choose k} J_{-n+2k}(z).
\end{equation}
The expression for the lensed CMB temperature correlation function given in equation~\eqref{eq:lensed_corr_fct_non_linear} is exact for any PDF of the lensing excursion angle that is symmetric about zero and whose moments of different components are uncorrelated, which we have shown to be valid in the deflection field obtained from the Millennium Simulation (cf. Section~\ref{subsec:moments}).

\section{Numerics}\label{sec:numerics}

\subsection{Truncation}\label{subsec:truncation}

For explicitly computing the lensed CMB temperature power spectrum via equation~\eqref{eq:lensed_corr_fct_non_linear} one has to truncate the expansion of the characteristic function at a certain order \(n\). To find a reasonable value for \(n\)  we resorted to the approximation of a purely Gaussian distributed lensing excursion angle and verified the performance of the series expansion in comparison to the lensing method of \citet{2005PhRvD..71j3010C} described in Section~\ref{subsec:lensing_ex_angle} and numerically implemented in CAMB\footnote{http://camb.info/}. Successively increasing the order \(n\) taken into account in the series expansion we determined that \(n\) for which both lensing methods work equivally well. We confirmed that in the case of a pure Gaussian lensing excursion angle PDF for \(n=3\) almost perfect agreement between both lensing methods can be achieved. The deviations are largest on very small scales but do not exceed \(\mathcal O ( 10^{-4})\).
Thus, for the numerical implementation of our lensing method we truncated the series expansion in equation~\eqref{eq:lensed_corr_fct_non_linear} at \(n=3\), i.e. we included the sixth moments in the series expansion. Explicit expressions for \(n=3\) are given in Appendix~\ref{sec:explicit_expressions}.

\subsection{Moments}\label{subsec:moments}

Aiming at the influence of the non-Gaussian features in the lensing excursion angle PDF on the lensed CMB power spectra we cannot compute the moments needed for its calculation analytically using the power spectrum approach developed by \citet{1994ApJ...436..509S, 1996ApJ...463....1S} (cf. Section~\ref{subsec:lensing_ex_angle}). This approach is well suited for describing gravitational light deflection by the intervening large scale structure but since it is based on linear perturbation theory it does not account for coherent lensing scatter by individual massive haloes, which gives rise to the exponential wings in the lensing excursion angle PDF (cf. Section~\ref{subsec:lensing_ex_angle}). Therefore, we either have to resort to numerical simulations or to use the halo model of large scale structure reviewed by \citet{2002PhR...372....1C}.

In this work the moments needed for the computation of the lensed CMB power spectra are obtained from the lensing deflection field constructed by \citet{2008MNRAS.388.1618C} which is based on the Millennium Simulation (MS) \citep{2006Natur.440.1137S}. Being an all-sky map the deflection field is given as angular gradient of the lensing potential
\begin{equation}
 \balpha(\mathbf{\hat{n}}) = \left( \mathbf{\hat{e}}_\theta \frac{\partial}{\partial \theta} 
	+ \mathbf{\hat{e}}_\phi \frac{1}{\sin\phi}\frac{\partial}{\partial \phi} \right) \psi(\mathbf{\hat{n}})
	= \alpha_\theta (\mathbf{\hat{n}}) \mathbf{\hat{e}}_\theta + \alpha_\phi(\mathbf{\hat{n}})\mathbf{\hat{e}}_\phi.
 \label{eq:spherical_gradient_of_lensing_potential}
\end{equation}
Since the non-Gaussian features of the lensing excursion angle PDF are only substantial for intrinsic light ray separations smaller than one degree (cf. Section~\ref{subsec:lensing_ex_angle}), i.e on scales where the curvature of the sky is negligible, we approximate
\begin{equation}
 \bdelta \balpha(r) = \balpha(\mathbf x) - \balpha (\mathbf x') \approx \bar{\alpha} (\mathbf{\hat{n}}) - \bar{\alpha} (\mathbf{\hat{n}}') = \bdelta \bar{\balpha} (\beta)
	\quad \textnormal{with} \quad r \approx \beta \quad \textnormal{and} \quad \mathbf{\hat{n}} \cdot \mathbf{\hat{n}}' = \cos \beta,
 \label{eq:approx_flat_full_sky}
\end{equation}
where the bars indicate the basis defined by the geodesic connecting \(\mathbf{\hat{n}}\) and \(\mathbf{\hat{n}}'\).
The error of this approximation is slightly increasing with the light ray separation and finally reaches \(\sim 2\%\) for two rays intrinsically separated by one degree.

\subsection{Correlation coefficients}\label{subsec:corr_coeff}

The expression for the lensed correlation function~\eqref{eq:lensed_corr_fct_non_linear} assumes that different moments of different components of the excursion angle are statistically independent. To show that this is indeed the case we compute the correlation coefficient defined by
\begin{equation}
 \rho(X,Y) = \frac{\left\langle (X-\left\langle X \right\rangle )(Y - \left\langle Y \right\rangle ) \right\rangle }
		{\sqrt{\left\langle (X - \left\langle X \right\rangle )^2 \right\rangle } \sqrt{\left\langle (Y - \left\langle Y \right\rangle )^2 \right\rangle }},
		\qquad X,Y \; \in \left\lbrace \left. \delta\alpha_{\theta,\phi}^n \right|n = 1,...,5 \right\rbrace
\end{equation}
for all relevant combinations of \(X\) and \(Y\). All \(\rho(X,Y)\) are compatible with zero at the level of \(10^{-3}\), justifying the assumption of statistical independence.

\subsection{Non-Gaussianity}\label{subsec:non_gaussianity}

\begin{figure}
 \centering
 \includegraphics[height=6.0cm,width=8.6cm]{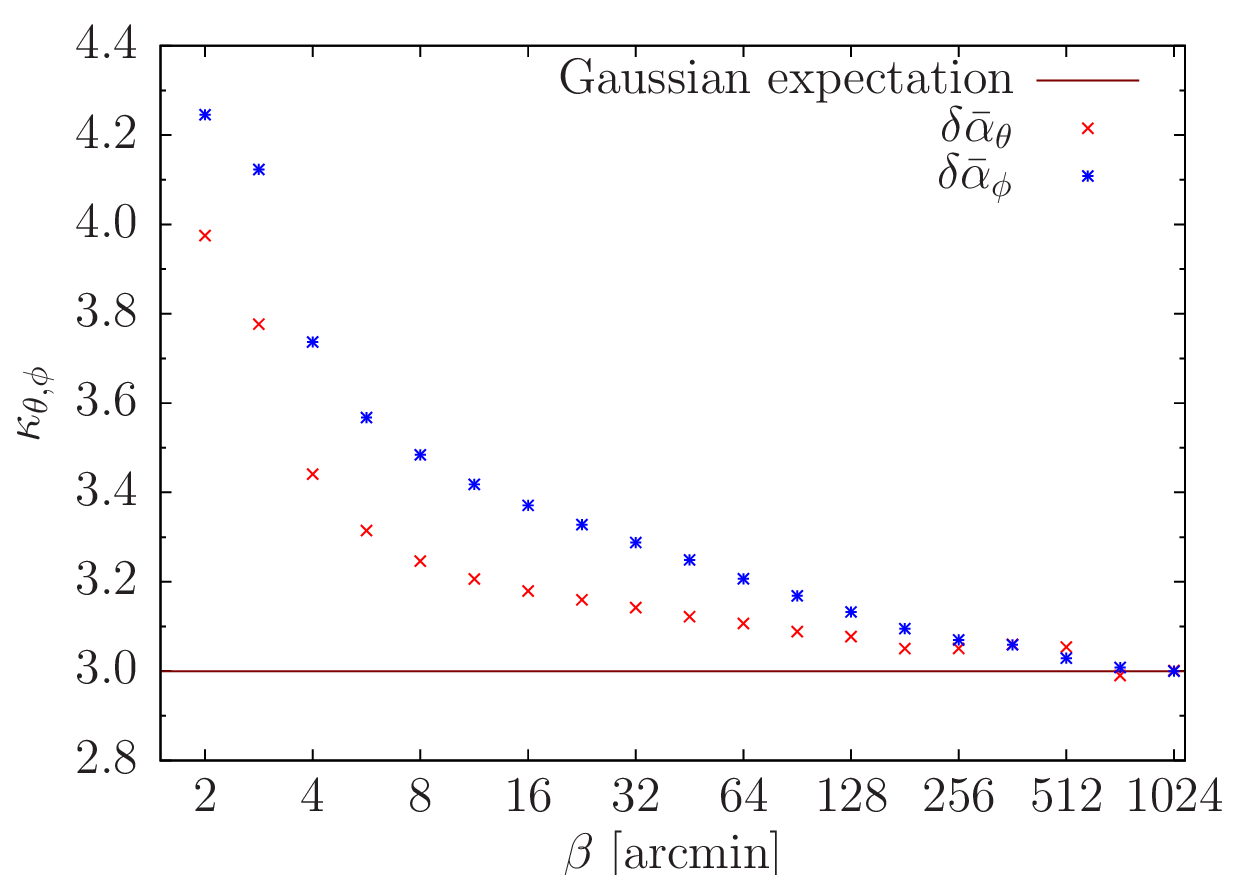}
 \hfil
 \includegraphics[height=6.0cm,width=8.6cm]{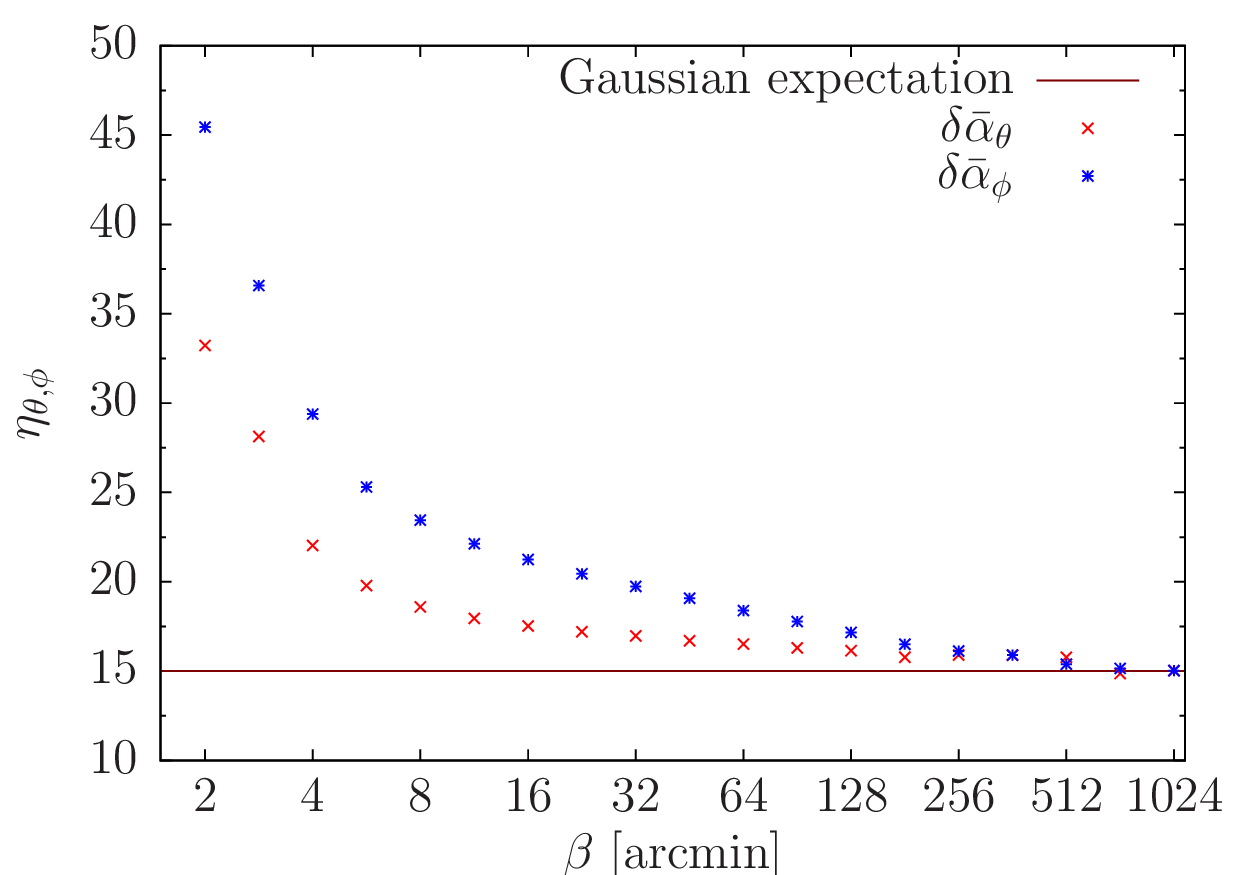}
 \caption{Quantifying the non-Gaussianity of the PDFs of the lensing excursion angle components, obtained from the MS deflection field, by the deviation of the fourth (left panel) and sixth (right panel) moment, respectively, from the Gaussian case, i.e. \(\kappa_{\theta,\phi} \equiv \left\langle \delta\bar\alpha_{\theta,\phi}^4 \right\rangle/\left\langle \delta\bar\alpha_{\theta,\phi}^2 \right\rangle^2\) and \(\eta_{\theta,\phi} \equiv \left\langle \delta\bar\alpha_{\theta,\phi}^6 \right\rangle / \left\langle \delta\bar\alpha_{\theta,\phi}^2 \right\rangle^3\). The deviations from the Gaussian expectation (\(\kappa_{\theta,\phi}\equiv3\), \(\eta_{\theta,\phi}\equiv15\)) decrease rapidly with increasing intrinsic light ray separation \(\beta\). For separations larger than one degree there are only small amounts of non-Gaussianity left.}
 \label{fig:kurtosen}
\end{figure}

Equation~\eqref{eq:lensed_corr_fct_non_linear} reveals that all information about the non-Gaussianity of the lensing excursion angle PDF is carried by its moments. It is therefore natural to quantify the amount of non-Gaussianity of the excursion angle PDF via comparing its moments with those of a Gaussian PDF. The moments of the latter can all be expressed in terms of the variance:
\begin{equation}
 \left\langle X^{2n} \right\rangle = (2n-1)!!\left\langle X^2 \right\rangle^n \quad \textnormal{with} \quad (2n-1)!! = (2n-1) \cdot (2n-2) \cdot ... \cdot 5 \cdot 3 \cdot 1.
\end{equation}
Thus 
\begin{equation}
 \kappa \equiv \frac{\left\langle X^4 \right\rangle}{\left\langle X^2 \right\rangle^2} \qquad \textnormal{and} \qquad 
	\eta \equiv \frac{\left\langle X^6 \right\rangle }{\left\langle X^2 \right\rangle^3}
 \label{eq:kappa_eta}
\end{equation}
or rather their deviation from 3 and 15 carry information about the amount of non-Gaussianity present in the fourth and sixth moment of any symmetric PDF.
Consequently, in order to ensure to capture only the effects on the lensed power spectrum which arise from the non-Gaussianity of the lensing excursion angle PDF it is recommended not to use directly the moments of the excursion angle components but to use instead \(\kappa_{\theta,\phi}\) and \(\eta_{\theta,\phi}\) defined by 
\begin{equation}
 \kappa_{\theta,\phi}(\beta) \equiv \frac{\left\langle \delta\bar\alpha_{\theta,\phi}^4 (\beta)\right\rangle}{\left\langle \delta\bar\alpha_{\theta,\phi}^2(\beta) \right\rangle^2}
	\qquad \textnormal{and} \qquad 
	\eta_{\theta,\phi}(\beta) \equiv \frac{\left\langle \delta\bar\alpha_{\theta,\phi}^6(\beta) \right\rangle }{\left\langle \delta\bar\alpha_{\theta,\phi}^2(\beta) \right\rangle^3}
 \label{eq:kappa_eta_for_l_e_a}
\end{equation}
in analogy to equation~\eqref{eq:kappa_eta}.
Thus we used
\begin{equation}
 \left\langle \delta{\alpha}_{x}^2(r) \right\rangle = \sigma^2 (r) + \cgltwo(r), \quad
 \left\langle \delta{\alpha}_{x}^4(r) \right\rangle = \kappa_{\theta}(r) \left\langle \delta{\alpha}_{x}^2(r) \right\rangle^2 
  \quad \textnormal{and} \quad
 \left\langle \delta{\alpha}_{x}^6(r) \right\rangle = \eta_{\theta}(r) \left\langle \delta{\alpha}_{x}^2(r) \right\rangle^3
 \label{eq:actually_used_moments_x}
\end{equation}
and
\begin{equation}
 \left\langle \delta{\alpha}_{y}^2(r) \right\rangle = \sigma^2 (r) - \cgltwo(r), \quad
 \left\langle \delta{\alpha}_{y}^4(r) \right\rangle = \kappa_{\phi}(r) \left\langle \delta{\alpha}_{y}^2(r) \right\rangle^2
 \quad \textnormal{and} \quad
 \left\langle \delta{\alpha}_{y}^6(r) \right\rangle = \eta_{\phi}(r) \left\langle \delta{\alpha}_{y}^2(r) \right\rangle^3,
 \label{eq:actually_used_moments_y}
\end{equation}
respectively, for our actual computation of the lensed correlation function accounting for the excess of small scale power of the lensing potential due to nonlinear growth via semianalytical corrections from the HALOFIT model \citep[cf.][]{2008MNRAS.388.1618C}.
For computing \(\kappa_{\theta,\phi}\) and \(\eta_{\theta,\phi}\) from the CMB deflection field of the MS we sampled pairs of pixels with fixed angular separation and computed their differences in the local coordinate system defined by the connecting geodesic. The rotation angle needed for the transion from the coordinate system used by the MS (cf. equation~\ref{eq:spherical_gradient_of_lensing_potential}) to the geodesic basis can be readily obtained from identities of spherical triangles. From these samples we estimated the first three even moments of the lensing excursion angle components and computed \(\kappa_{\theta,\phi}\) and \(\eta_{\theta,\phi}\). They are shown in Figure~\ref{fig:kurtosen}. Note that the sampling errors are negligible and are not shown.
The non-Gaussian features are more pronounced in the distribution of the \(\phi\)-component of the lensing excursion angle. The deviations from the Gaussian expectation are larger in case of \(\eta_{\theta,\phi}\) than for \(\kappa_{\theta,\phi}\) as expected since the PDFs are broadened due the coherent scattering by nonlinear structures (cf. Section~\ref{subsec:lensing_ex_angle}). Furthermore Figure~\ref{fig:kurtosen} reveals that there is still a small amount of non-Gaussianity for light ray separations larger than one degree. The Gaussian expectation is not reached until \(\beta  \sim \)~20\degr. For such large light ray separations, however, the approximation \( \bdelta\bar{\balpha}(\beta) \approx \bdelta\balpha(r)\) given in equation~\eqref{eq:approx_flat_full_sky} is not longer valid. Since the error from extending this approximation up to separations larger than one degree would fairly exceed the one resulting from neglecting the small amount of non-Gaussianity present for {\(\beta > \)~1\degr} the latter is omitted in the remainder of this work.

\subsection{Results}\label{subsec:results}

As seen in Section~\ref{subsec:moments}, for intrinsic light ray separations up to one degree the lensing excursion angle can be well approximated by the difference of two deflection angles defined in the geodesic basis. In the regime of larger separations the non-Gaussianity of the lensing excursion angle PDFs is weak and we confirmed that in this regime the Gaussian approximation is very good. Thus, it is natural to establish the following computation scheme for the lensed correlation function: For intrinsic light ray separations less than one degree one uses formula~\eqref{eq:lensed_corr_fct_non_linear} together with the moments computed via equations~\eqref{eq:actually_used_moments_x} and~\eqref{eq:actually_used_moments_y}. For larger separations one uses equation~\eqref{eq:lensed_correlation_function} together with equation~\eqref{eq:gaussian_char_fct}, which is valid for Gaussian distributed excursion angles. The same computation scheme can be applied to the lensed CMB polarization power spectra by using the corresponding equations given in the appendix. 

\begin{figure}
 \centering
 \includegraphics[height=5.5cm,width=8.6cm]{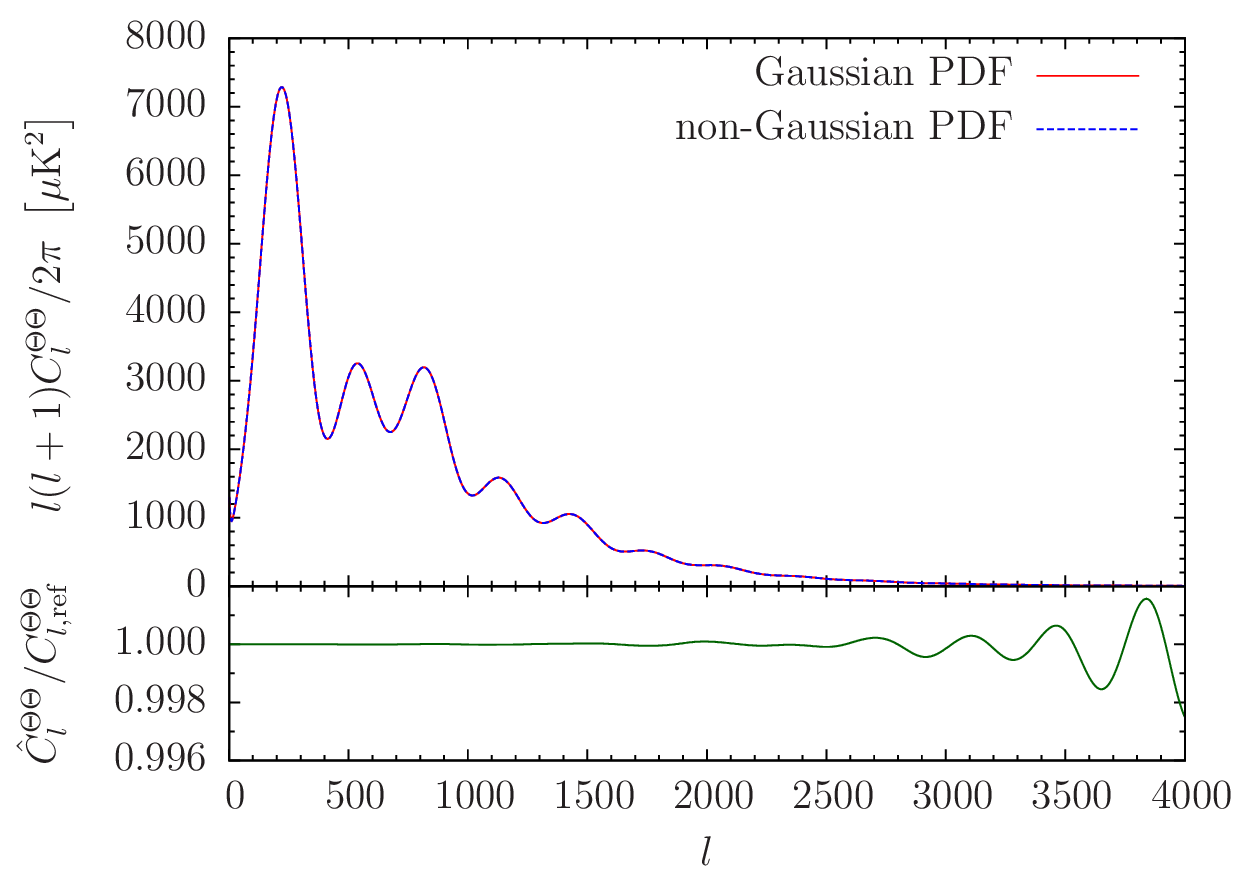}
 \hfil
 \includegraphics[height=5.5cm,width=8.6cm]{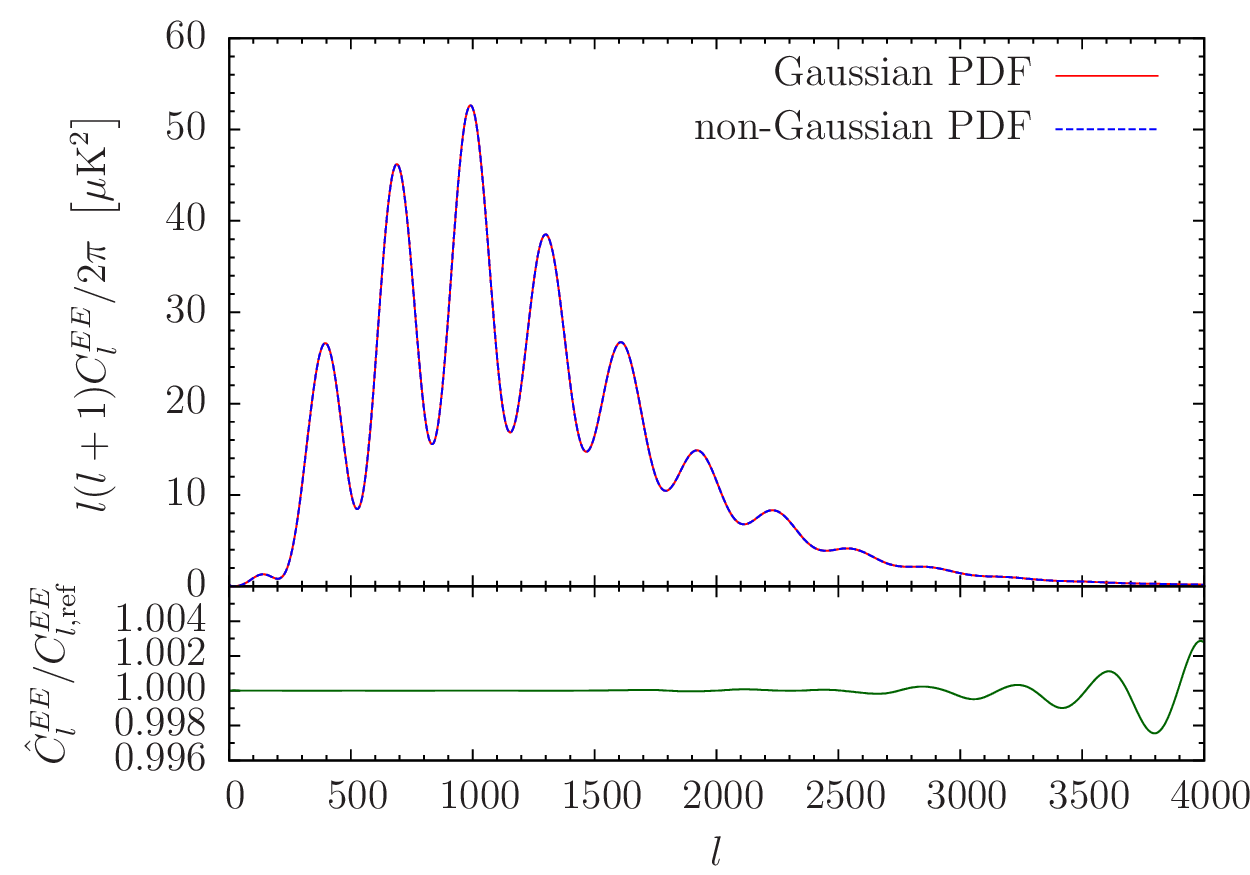}
 \\
 \includegraphics[height=5.5cm,width=8.6cm]{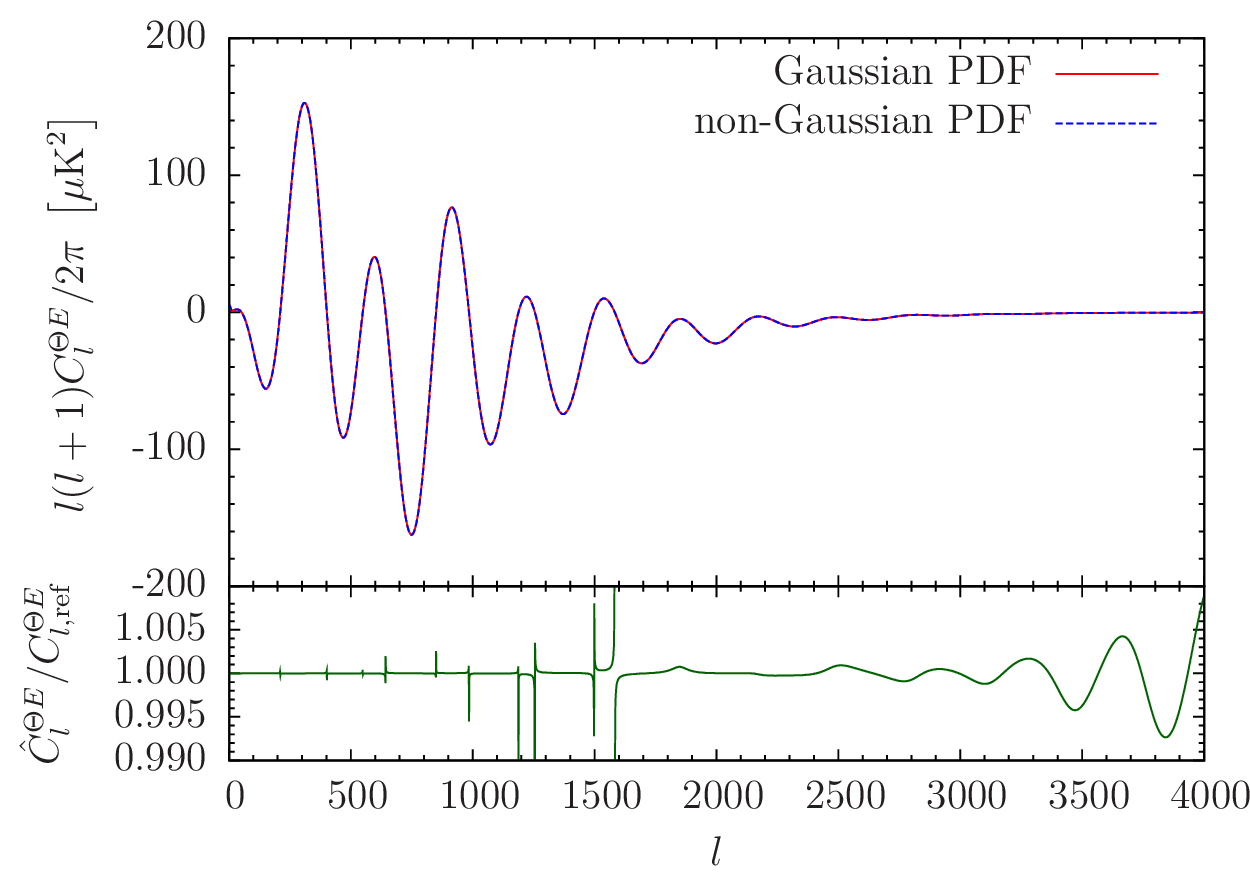}
 \hfil
 \includegraphics[height=5.5cm,width=8.6cm]{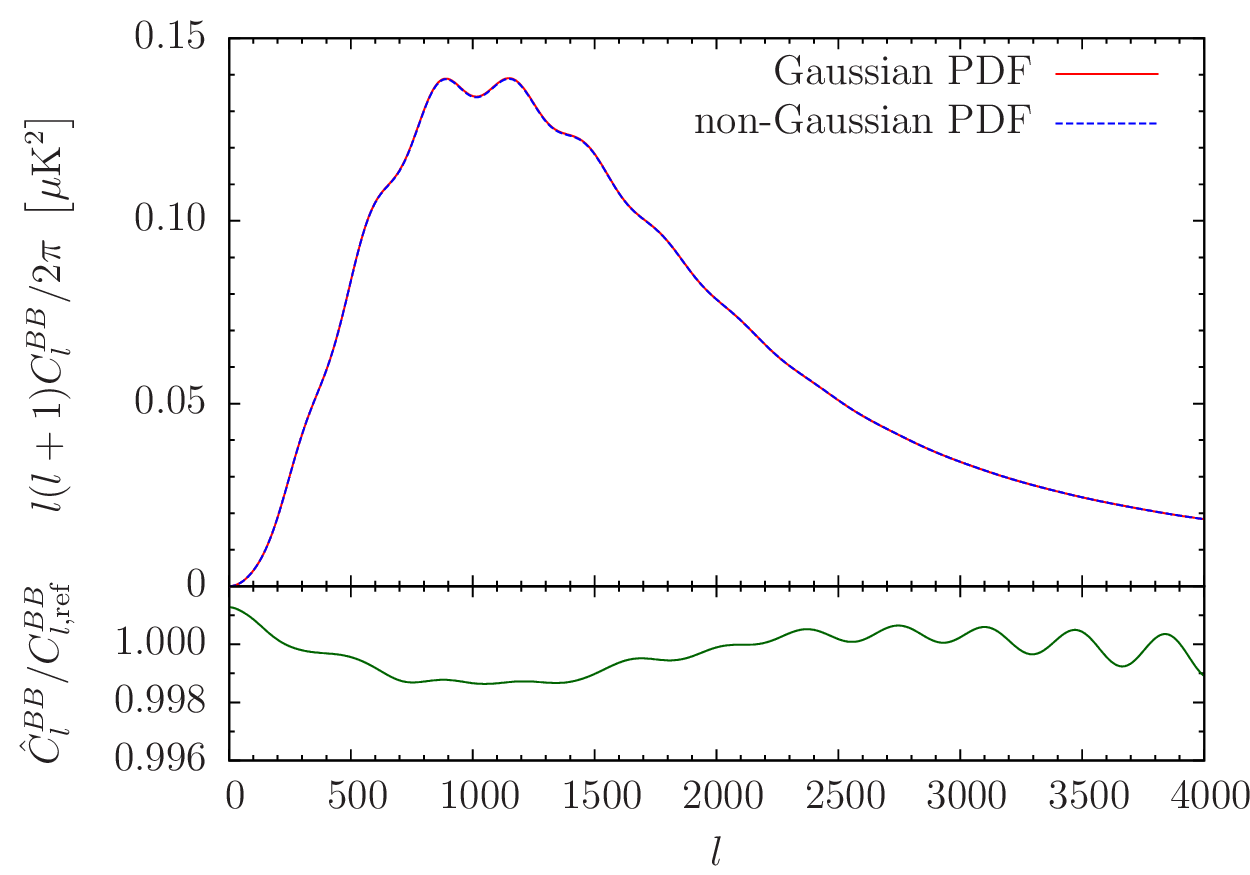}
 \caption{	Lensed CMB power spectra computed via the method described in the text accounting for the non-Gaussian features of the lensing excursion angle PDF discussed in
		Section~\ref{subsec:lensing_ex_angle} (blue dashed curves). The reference power spectra (solid red curves) were computed by the same method but assume a Gaussian
		 PDF of the lensing excursion angle. The ratio of the corresponding spectra is shown in the lower part of each panel (solid green lines).
	}
 \label{fig:lensed_power_spectra}
\end{figure}

The influence of a non-Gaussian PDF on the lensed CMB power spectra can now be quantified by computing two sets of power spectra \(\hat{C}_l^{XY}\) and \(\hat{C}_{l,\mathrm{ref}}^{XY}\). For both sets the computation scheme for the lensed correlation functions described above is used but in case of the reference spectra we use Gaussian moments on all scales, i.e. setting \(\kappa_{\theta,\phi}(r) \equiv 3\) and \(\eta_{\theta,\phi}(r) \equiv 15\). Figure~\ref{fig:lensed_power_spectra}  shows both the lensed and reference CMB power spectra. In the lower half of each panel the ratio between the corresponding spectra is depicted.

These ratios show that the influence of the non-Gaussianity in the lensing excursion angle distribution function is marginal. The differences caused by the exponential wings of the PDF are just several per mile on very small scales \((\Delta \theta \lesssim 4', \, l \ga 2500) \). On large and intermediate scales all lensed power spectra but the \(B\)-modes are almost unaffected. The lensed \(B\)-modes, however, are altered on all scales, reflecting the fact that in the cosmological model assumed in this work primordial gravitational waves are absent and thus polarization of the \(B\)-type is completely lensing induced.

Taking into account that neglecting the curvature of the sky, as we did in this work, already causes an error in the lensed CMB power spectra at the 0.3\% --1.0\% level \citep{2005PhRvD..71j3010C} we conclude that the influence of the non-Gaussian part of the lensing excursion angle PDF on the lensed CMB spectra is far from being observable and can be safely neglected. Furthermore, it is very likely that uncertainties in the recombination history and contamination from various late-time secondaries affect the lensed power spectra in an even stronger way \citep{2006PhR...429....1L}.

Our results have been anticipated by the purely numerical work of \citet{2009MNRAS.396..668C}. Using the MS they constructed lensed CMB temperature and polarization maps and computed the corresponding power spectra. They agree well with the ones obtained from CAMB including nonlinear corrections to the lensing potential from the HALOFIT model. Hence, \citet{2009MNRAS.396..668C} increased the variance of the lensing excursion angle but did not drop the assumption of a Gaussian PDF \citep[cf.][]{2005PhRvD..71h3008L}.

Our analytical approach, however, takes the non-Gaussianity of the lensing excursion angle PDF explicitly into account revealing that its influence on the lensed CMB spectra is weak and therefore the assumption of a pure Gaussian lensing excursion angle PDF is well applicable. The weakness of the impact of the non-Gaussian features is due to the fact that they are only prominent on small scales (see Figure~\ref{fig:kurtosen}), while lensing smooths the CMB signal over a large angular range.

\section{Summary}\label{sect_summary}
The topic of this paper is a derivation of the fluctuation statistics of the lensed CMB temperature and polarisation, taking non-Gaussian features of the lensing deflection angle distribution into account. 
\begin{enumerate}
\item{Starting point of including non-Gaussian distributions of the deflection angle is the expansion of the characteristic function, which enters the computation of the lensed CMB spectra, into a a series in terms of its moments. For computing lensed CMB spectra with non-Gaussian deflection angle fields we provide a set of analytical expressions, and expand the expressions up to the sixth-order moment of the deflection angle distribution.}
\item{Simulated deflection maps \citep[provided by ][]{2009MNRAS.396..668C} show considerable amounts of non-Gaussianity on small scales below a degree, which we quantified with the fourth and sixth moment. On angular scales of an arcminute, they exceed the Gaussian expectation by a factor of 1.5 and 3, respectively, and drop close to their fiducial, Gaussian values close to one degree.}
\item{We show that deviations in the angular spectra relative to those derived assuming Gaussian statistics are most important on small angular scales, but remain below the percent level, confirming that the Gaussian approximation is very good. In particular we confirm that errors introduced into the spectra by non-Gaussian deflection angle statistics are smaller than those caused by other secondary anisotropies.}
\item{One can give a simple physical argument, why the impact of the non-Gaussian lensing excursion angle PDF is so weak: On the subdegree scale, where the non-Gaussianity is considerable, the CMB spectra are almost featureless. As lensing cannot generate features in a featureless CMB \citep[see][]{2000PhRvD..62d3007H,2006PhR...429....1L} the influence of the non-Gaussianity of the lensing excursion angle, which itself is only substantial on subdegree scales, is very weak.}
\item{Estimates of cosmological parameters and weak lensing reconstructions are not seriously impeded by non-Gaussianities in the deflection angle distribution, given the small differences relative to spectra derived with a Gaussian approximation.
Furthermore, the differences in the spectra involving the temperature and \(E\)-type polarization are substantial only at high multipole order where the signal-to-noise ratio is small and where the spectra do not possess strong parameter constraining power due to Silk-damping.}
\end{enumerate}
The formalism presented here can be applied to investigating primordial non-Gaussianities by CMB lensing e.g. in terms of the $f_\mathrm{NL}$-model or the $\chi^2$-model, for which the higher-order moments are directly calculable. It would be interesting to see if lensed CMB spectra are significantly distorted in these cosmological models, because they inherently are able to provided stronger non-Gaussian features on larger angular scales compared to non-Gaussianities generated by nonlinear structure formation.

\section*{Acknowledgements}
We would like to thank Matthias Bartelmann for valuable comments, and Carmelita Carbone for providing the lensing deflection map. For some of our numerical results we used routines of the HEALPix package \citep{2005ApJ...622..759G}. We are grateful to Martin Reinecke for his help with these routines. BMS's work is supported by the German Research Foundation (DFG) within the framework of the excellence initiative through the Heidelberg Graduate School of Fundamental Physics.

\bibliography{bibtex/aamnem,bibtex/references}
\bibliographystyle{mn2e}

\appendix

\section{Polarization}\label{sec:polarization}

\subsection{Correlation functions and power spectra}\label{subsec:pol_corr_fct_and_pow_spec}

The polarization of the CMB is described by the Stokes parameters \(Q\) and \(U\) constituting the spin-2 polarization field \(P = Q +\mathrm i U\), whose decomposition in gradient-like \(E\)- and curl-like \(B\)-modes reads (in the limit of a flat sky)
\begin{equation}
 P(\mathbf x ) = - \int \frac{\dd^2 l}{2\pi} (E(\mathbf l) - \mathrm i B (\mathbf l) ) e^{-2\mathrm i \phi} e^{\mathrm i \mathbf l \cdot \mathbf x}
\end{equation}
\citep{2005PhRvD..71j3010C}.
Defining the correlation functions involving \(P\) in the same local coordinate system as before, i.e. with the \(x\)-axis  adapted to the vector connecting \(\mathbf x\) and \(\mathbf x'\), we have
\begin{equation}
 \xi_+(r) \equiv \left\langle P^*(\mathbf x) 
		P(\mathbf x') \right\rangle,
	\quad
 \xi_-(r) \equiv \left\langle P(\mathbf x) 
		P(\mathbf x') \right\rangle
	\quad \textnormal{and} \quad
 \xi_\times(r) \equiv \left\langle \Theta(\mathbf x)
		 P(\mathbf x') \right\rangle
\end{equation}
\citep{2005PhRvD..71j3010C}.
The corresponding power spectra are then calculated via
\begin{equation}
 C_l^{EE} + C_l^{BB} = 2\pi \int r\dd r \; J_0(lr) \xi_+(r),
 \quad
 C_l^{EE} - C_l^{BB} = 2\pi \int r\dd r \; J_4(lr) \xi_-(r)
 \quad \textnormal{and} \quad
 C_l^{\Theta E} = 2\pi \int r\dd r \; J_2(lr) \xi_\times(r)
\end{equation}
\citep{2006PhR...429....1L}.

\subsection{Lensed correlation functions}\label{subsec:pol_lensed_corr_fct}

The lensed correlation functions involving the polarization field valid for a general distribution function of the lensing excursion angle can be derived in complete analogy to Section~\ref{subsec:non_gaussian_pdf}. The expression for \(\hat\xi_+(r)\) is identical with equation~\eqref{eq:lensed_corr_fct_non_linear} if one replaces \(C^{\ttt}_l\) by \(C^{EE}_l + C^{BB}_l\):
\begin{eqnarray}
 \hat\xi_+(r) &=& \int \frac{\dd^2l}{(2\pi)^2} \left(C^{EE}_l + C^{BB}_l\right) e^{\mathrm i \mathbf l \cdot \mathbf r} 
			\left\langle \sum_{n=0}^{\infty} \frac{\left( \mathrm i \mathbf l \cdot \bdelta \balpha\right)^n}{n!}\right\rangle
 \nonumber\\
 &=& \frac{1}{2\pi} \int l\dd l \; \left(C^{EE}_l + C^{BB}_l\right)
			\sum_{n=0}^\infty \sum_{k=0}^{n} \sum_{q=0}^{n-k} \sum_{r=0}^{2(k+q)}
			\frac{(-1)^{n+k+r}}{(2n)!4^{k+q}} {2n \choose 2k} {n-k \choose q} {2(k+q) \choose r}
			J_{2(k+q-r)}(lr)
	\left\langle \delta\alpha_x^{2k} \right\rangle \left\langle \delta\alpha_y^{2(n-k)}\right\rangle.
 \label{eq:xi_plus_nonlinear}
\end{eqnarray}
Some additional effort, however, has to be put in the computation of \(\hat\xi_-(r)\) and \(\hat\xi_\times(r)\), since here the additional factors of \(e^{-2\mathrm i \phi}\) in the spin 2 polarization do not cancel. After a somewhat lengthy but straightforward calculation we find
\begin{eqnarray}
	\hat\xi_-(r)
	& = &
	\int \frac{\dd^2 l }{(2\pi)^2} \left( C_l^{EE} - C_l^{BB} \right) \cos4\phi \:
	e^{\mathrm i \mathbf l \cdot \mathbf r}
	\left\langle \sum_{n=0}^{\infty} \frac{\left( \mathrm i \mathbf l \cdot \bdelta \balpha\right)^n}{n!}\right\rangle
	\nonumber\\
	& = &
	\frac{1}{2\pi} \int l \dd l \; \left( C_l^{EE} - C_l^{BB} \right)
	\sum_{n=0}^{\infty} \sum_{k=0}^{n} \sum_{q=0}^{n-k} \frac{(-1)^{n+k}}{(2n)!4^{k+q}} {2n\choose 2k}{n-k \choose q}
	\left\langle \delta\alpha_x^{2k} \right\rangle \left\langle \delta\alpha_y^{2(n-k)}\right\rangle
	\nonumber\\
	&&
	\qquad \cdot \: \left[
		8 \sum_{r=0}^{2(k+q+2)}(-1)^r {2(k+q+2) \choose r} J_{2(k+q+2-r)}(lr)
		-8 \sum_{r=0}^{2(k+q+1)}(-1)^r {2(k+q+1) \choose r} J_{2(k+q+1-r)}(lr)
	\right.
	\nonumber\\
	&&
	\left.
		\qquad \qquad \:+\sum_{r=0}^{2(k+q)}(-1)^r {2(k+q) \choose r} J_{2(k+q-r)}(lr)
	\right]
 \label{eq:xi_minus_nonlinear}
\end{eqnarray}
and
\begin{eqnarray}
	\hat\xi_\times (r)
	& = &
	 - \int \frac{\dd^2 l }{(2\pi)^2} C_l^{\Theta E} \cos2\phi \:
	e^{\mathrm i \mathbf l \cdot \mathbf r}
	\left\langle \sum_{n=0}^{\infty} \frac{\left( \mathrm i \mathbf l \cdot \bdelta \balpha\right)^n}{n!}\right\rangle
	\\
	& = &
	\frac{1}{2\pi} \int l\dd l\; C_l^{\Theta E} \sum_{n=0}^\infty \sum_{k=0}^{n} \sum_{q=0}^{n-k}
	\frac{(-1)^{k+n}}{(2n)! 4^{k+q}} {2n \choose 2k} {n-k \choose q}
	\left\langle \delta\alpha_x^{2k} \right\rangle \left\langle \delta\alpha_y^{2(n-k)}\right\rangle 
	\nonumber\\
	& &
	\qquad \cdot \: \left[ 2 \sum_{r=0}^{2(k+q+1)} (-1)^r {2(k+q+1) \choose r} J_{2(k+q+1-r)}(lr) 
	+ \sum_{r=0}^{2(k+q)} (-1)^r {2(k+q) \choose r} J_{2(k+q-r)}(lr) \right].
 \label{eq:xi_X_nonlinear}
\end{eqnarray}

\section{Explicit expressions}\label{sec:explicit_expressions}

Here we give explicit expressions for the lensed correlation functions used for our numerical calculations. Truncating at \(n = 3\) in the equations~\eqref{eq:lensed_corr_fct_non_linear}, \eqref{eq:xi_plus_nonlinear}, \eqref{eq:xi_minus_nonlinear} and \eqref{eq:xi_X_nonlinear}, we obtain:
\begin{eqnarray}
	\hat\xi (r)
	& = & 
	\frac{1}{2\pi}\int l\mathrm dl\; C_{l}^{\Theta\Theta} \biggl( J_0(lr)
	-\frac{1}{4}l^2 \left[\left(\dax{2}+\dayy{2}  \right)J_0(lr) 
		- \left(\dax{2} -\dayy{2} \right)J_2(lr)\right]\biggr.
	\nonumber\\
 	& & \qquad + \frac{1}{24}l^4 \left[\frac{1}{8}\left(\dax{4} + \dayy{4}\right)\left(3J_0(lr)+J_4(lr)\right)
	-\frac{1}{2}\left(\dax{4}-\dayy{4}\right)J_2(lr) \right.
	\nonumber\\
 	& &
	\qquad\qquad\qquad\quad \left.+\frac{3}{4} \dax{2}\dayy{2}\left(J_0(lr)-J_4(lr)\right)\right]
	\nonumber\\
	& &
	\qquad - \frac{1}{720}l^6 \left[\frac{1}{16} \left(\dax{6}+\dayy{6}\right) \left(5J_0(lr) +3J_4(lr)\right)
	+ \frac{1}{32}\left(\dax{6}-\dayy{6}\right)\left(-15J_2(lr)-J_6(lr)\right)\right.
	\nonumber\\
	& &
	\qquad\qquad\qquad\quad 
	+\frac{15}{32}\dax{2}\dayy{4}\left(2J_0(lr)+J_2(lr)-2J_4(lr)-J_6(lr)\right)
	\nonumber\\
	& &
	\qquad\qquad\qquad\qquad
	\left.+\frac{15}{32}\dax{4}\dayy{2}\left(2J_0(lr)-J_2(lr)-2J_4(lr)+J_6(lr)\right)\right] 
	\biggl. +  \; ... \;\biggr),
\end{eqnarray}
\begin{eqnarray}
	\hat\xi_+(r) 
	& = &
	\frac{1}{2\pi}\int l\mathrm dl\; \left(C^{EE}_l+C^{BB}_l\right) \biggl( J_0(lr)
	-\frac{1}{4}l^2 \left[\left(\dax{2}+\dayy{2}  \right)J_0(lr)
	- \left(\dax{2} -\dayy{2} \right)J_2(lr)\right]\biggr.
	\nonumber\\
	& &
	\qquad + \frac{1}{24}l^4 \left[\frac{1}{8}\left(\dax{4} + \dayy{4}\right)\left(3J_0(lr)+J_4(lr)\right)
	-\frac{1}{2}\left(\dax{4}-\dayy{4}\right)J_2(lr) \right.
	\nonumber\\
	& &
	\qquad\qquad\qquad\quad \left.+\frac{3}{4} \dax{2}\dayy{2}\left(J_0(lr)-J_4(lr)\right)\right]
	\nonumber\\
	& &
	\qquad - \frac{1}{720}l^6 \left[\frac{1}{16} \left(\dax{6}+\dayy{6}\right) \left(5J_0(lr) +3J_4(lr)\right)
	+ \frac{1}{32}\left(\dax{6}-\dayy{6}\right)\left(-15J_2(lr)-J_6(lr)\right)\right.
	\nonumber\\
	& &
	\qquad\qquad\qquad\quad
	+\frac{15}{32}\dax{2}\dayy{4}\left(2J_0(lr)+J_2(lr)-2J_4(lr)-J_6(lr)\right)
	\nonumber\\
 	& &
	\qquad\qquad\qquad\qquad
	\left.+\frac{15}{32}\dax{4}\dayy{2}\left(2J_0(lr)-J_2(lr)-2J_4(lr)+J_6(lr)\right)\right] 
	\biggl. + \; ... \;\biggr),
\end{eqnarray}
\begin{eqnarray}
	\hat\xi_-(r) 
	& = &
	\frac{1}{2\pi}\int l\mathrm dl\; \left(C^{EE}_l - C^{BB}_l\right) \biggl( J_4(lr)
	-\frac{1}{4}l^2 \left[\left(\dax{2}+\dayy{2} \right)J_4(lr) 
	- \frac{1}{2}\left(\dax{2} -\dayy{2} \right)\left(J_2(lr)+J_6(lr)\right)\right]\biggr.
	\nonumber\\
	& &
	\qquad + \frac{1}{24}l^4 \left[\frac{1}{16}\left(\dax{4} +
	 \dayy{4}\right)\left(J_0(lr)+6J_4(lr)+J_8(lr)\right)
	-\frac{1}{4}\left(\dax{4}-\dayy{4}\right)\left(J_2(lr)+J_6(lr)\right) \right.
	\nonumber\\
	& &
	\qquad\qquad\qquad\quad \left.-\frac{3}{8} \dax{2}\dayy{2}\left(J_0(lr)-2J_4(lr)+J_8(lr)\right)\right]
	\nonumber\\
	& &
	\qquad - \frac{1}{720}l^6 \left[\frac{1}{32} \left(\dax{6}+\dayy{6}\right) \left(3J_0(lr)
	 +10J_4(lr)+3J_8(lr)\right)\right.
	\nonumber\\
 	& & 
	\qquad\qquad\qquad\quad -\frac{1}{64}\left(\dax{6}-\dayy{6}\right)
	\left(16J_2(lr)+15J_6(lr)+J_{10}(lr)\right)
	\nonumber\\
	& &
	\qquad\qquad\qquad\qquad\quad
	+\frac{15}{64}\dax{2}\dayy{4}\left(-2J_0(lr)+4J_4(lr)+J_6(lr)-2J_8(lr)-J_{10}(lr)\right) 
	\nonumber\\
	& &
	\qquad\qquad\qquad\qquad\qquad
	\left.+\frac{15}{64}\dax{4}\dayy{2}\left(-2J_0(lr)+4J_4(lr)-J_6(lr)-2J_8(lr)+J_{10}(lr)\right)\right]
	\biggl. + \; ... \;\biggr)
\end{eqnarray}
and
\begin{eqnarray}
	\hat\xi_\times(r)
	& = &
	\frac{1}{2\pi}\int l\mathrm dl\; C^{\Theta E}_l \biggl( J_2(lr)
	- \frac{1}{4}l^2 \left[\left(\dax{2}+\dayy{2}  \right)J_2(lr) 
	- \frac{1}{2}\left(\dax{2} -\dayy{2} \right)\left(J_0(lr)+J_4(lr)\right)\right]\biggr.
	\nonumber\\
	& & 
	\qquad - \frac{1}{24}l^4 \left[-\frac{1}{16}\left(\dax{4} +
	\dayy{4}\right)\left(7J_2(lr)+J_6(lr)\right)
	+\frac{1}{4}\left(\dax{4}-\dayy{4}\right)\left(J_0(lr)+J_4(lr)\right) \right.
	\nonumber\\
	&&
	\qquad\qquad\qquad\quad \left.+\frac{3}{8} \dax{2}\dayy{2}\left(-J_2(lr)+J_6(lr)\right)\right]
	\nonumber\\
	&&
	\qquad + \frac{1}{720}l^6 \left[-\frac{1}{32} \left(\dax{6}+\dayy{6}\right) \left(13J_2(lr)
	 +3J_6(lr)\right)\right.
	\nonumber\\
	&&
	\qquad\qquad\qquad\quad +\frac{1}{64}\left(\dax{6}-\dayy{6}\right)
	\left(15J_0(lr)+16J_4(lr)+J_{8}(lr)\right) 
	\nonumber\\
	&&
	\qquad\qquad\qquad\qquad\quad 
	+\frac{15}{64}\dax{2}\dayy{4}\left(-J_0(lr)-2J_2(lr)+2J_6(lr)+J_8(lr)\right) 
	\nonumber\\
	&& \qquad\qquad\qquad\qquad\qquad
	\left.+\frac{15}{64}\dax{4}\dayy{2}\left(J_0(lr)-2J_2(lr)+2J_6(lr)-J_8(lr)\right)\right] 
	\biggl. + \; ... \;\biggr).
\end{eqnarray}

\bsp

\label{lastpage}

\end{document}